\documentclass[aps,prd,showpacs,onecolumn,nofootinbib,superscriptaddress,amsmath,amssymb]{revtex4}
\usepackage{epsfig}
\usepackage{graphicx}
\usepackage{color}
\usepackage{mathrsfs}   
\usepackage{bm}   

\usepackage{caption}         
\usepackage{subcaption}    
\begin{document}
\title{Dissipative Callan-Harvey mechanism in 2+1 D Dirac system:
The fate of edge states along a domain wall}

\author{C. X. Zhang}
\affiliation{Institute for Theoretical Physics, Julius-Maximilians-Universit\"at W\"urzburg,
   97074 W\"urzburg, Germany}
\affiliation{W\"urzburg -Dresden Cluster of Excellence ct.qmat}
   
\author{M. Ulybyshev}
\affiliation{Institute for Theoretical Physics, Julius-Maximilians-Universit\"at W\"urzburg,
   97074 W\"urzburg, Germany}
 
\author{C. Northe}
\affiliation{Institute for Theoretical Physics, Julius-Maximilians-Universit\"at W\"urzburg,
   97074 W\"urzburg, Germany}
\affiliation{W\"urzburg -Dresden Cluster of Excellence ct.qmat}
\affiliation{Department of Physics, Ben-Gurion University of the Negev,
David Ben Gurion Boulevard 1, Be’er Sheva 84105, Israel}
  
\author{E. M. Hankiewicz}
\affiliation{Institute for Theoretical Physics, Julius-Maximilians-Universit\"at W\"urzburg,
   97074 W\"urzburg, Germany}
\affiliation{W\"urzburg -Dresden Cluster of Excellence ct.qmat}

\begin{abstract}
The Callan-Harvey mechanism in 2+1 D Jackiw-Rebbi model is revisited.
We analyzed Callan-Harvey anomaly inflow in the massive Chern insulator
(quantum anomalous Hall system) subject
to external electric field. In addition to the conventional current
flowing from the bulk to edge due to parity anomaly, we considered
the dissipation of the edge charge due to interaction with external
bosonic bath in 2+1 D and due to external bath of photons in 3+1D.
In the case of 2+1 D bosonic bath, we
found the new stationary state, which is defined by the balance
between Callan-Harvey current and the outgoing flow caused by the
dissipation processes.
In the case of 3+1 D photon bath, we found a critical electric field,
below which this balance state can be achieved, but above which
there is no such a balance. Furthermore, we estimated the 
photon-mediated transition rate between 2+1 D bulk and 1+1 D topological 
edge state of the order of one ns$^{-1}$  
at the room temperature.
\end{abstract}

\maketitle

\section{Introduction}
An anomaly in quantum field theory (or quantum anomaly) occurs when
a symmetry of the classical action is broken by quantum effects.
One of the most important quantum anomalies is the chiral anomaly, also known as
Adler-Bell-Jackiw anomaly \cite{Adler_1969,Bell_1969} or axial anomaly.
It is related to the breaking of the conservation law of an axial vector
current, which is associated with chiral symmetry, by quantum fluctuations.
In odd-dimensional
space-time, the chiral anomaly does not exist, and is replaced by the so-called parity anomaly:
if fermions are coupled to a gauge field,
parity symmetry is lost after quantization.
These quantum anomalies evoked great research interest in elementary particle physics
and in condensed matter physics.
The chiral anomaly is important to understand the pion decay into
two photons ($\pi \rightarrow \gamma\gamma$) and also the chiral magnetic effect
in Dirac materials\cite{Kharzeev_2014}. 
In contrast, the parity anomaly is essential in the quantum anomalous
Hall effect (QAHE) \cite{Haldane_1988}, 
which is defined as
quantized Hall conductivity in the absence of a magnetic field\cite{Liu+Zhang+Qi_2016}. 
In both scenarios, anomalies confirm the deviation from classical physics.
Thus, their presence adds to the long list of successes of quantum theory.

Parity anomaly and chiral anomaly show some certain connection when one considers a
finite-size fermionic system with boundaries \cite{Callan_1985}.
We take a cylinder-shaped bulk system in 2+1 D with two 1+1 D edges \cite{Laughlin_1981} as an example, and consider two scenarios to review such
a connection.
The first scenario is the work done by one of the authors \cite{Boettcher_2019}.
Due to the parity anomaly in the 2+1 D bulk, an out-of-surface magnetic field
pumps the charge to the bulk states \cite{Niemi_1983}, but the total charge density 
$n_{tot}=n_{bulk}+n_{edge}$
is constant and zero.
It demonstrates the Callan-Harvey mechanism
\cite{Callan_1985}: it is the edge states that compensate the 
charge deficit of bulk under the magnetic field \cite{Boettcher_2019}.
The second scenario is in the absence of
the magnetic field, but
in the presence of an electric field parallel to the edges,
which induces a Hall current in the bulk, perpendicular to the edge, due to the parity anomaly.
This bulk current pumps charge from one edge across the bulk to the other edge,
and the charge accumulates at the edges, which changes the
chemical potentials between them. 
This is another example demonstrating the Callan-Harvey mechanism
\cite{Callan_1985}\cite{Chandrasekharan_1994}:
From the viewpoint of the bulk, the current "stops" at the edge, which breaks
charge conservation. At the same time, the electric field generates charges
at the edge, because of the 1+1 D chiral anomaly.
One has to consider the two subsystems together; only then the charge conservation law holds for the whole system. Importantly this cancels the gauge anomaly \cite{Tong_string} that would otherwise occur.

Now one may ask the following questions:
What is the fate
of this surplus charge at the edge?
Will the charge accumulation be boundless?
We know that such an edge mode
propagates in a single direction,
and is protected by topology. Back-scattering
is forbidden, which makes the edge mode robust
to impurities \cite{Pashinsky+Goldstein_2020}.
However, the accumulation cannot happen infinitely;
when all the edge states are occupied, one expects relaxation to the bulk bands.
In reality, however, the edge states and the bulk states interact with each other.
One expects that if the edge chemical potential
is higher than the energy gap of the bulk,
say $\mu > m_0 v^2$, relaxation occurs: the electrons
at the high-energy (occupied) edge states
tend to relax into the low-energy (empty) bulk states,
and dissipate energy to the environment.
Such an interplay has been investigated in quantum Hall systems
\cite{Heinonen_1992}\cite{Lafont_2014}.

In the present work, we study the interplay between edge states and
bulk states in QAHE systems by introducing electron-photon interactions
\cite{Ulybyshev_2016}.
In addition to the edge-to-bulk relaxation process,
there is another excitation process transferring the charge from an edge state to the bulk. Even before the edge chemical potential exceeds the gap energy, i.e. $\mu < m_0 v^2$, the edge state can
be excited into bulk states by absorbing a photon from the thermal fluctuations.
Such an excitation is the
leading order contribution to the transition, pushing the
electrons to leave the edge.
Our present work will focus on such an excitation process
and we will calculate its rate using the Lindblad formalism.

The paper is organized as follows. In Sec.2, the
Jackiw-Rebbi model is introduced, and the Callan-Harvey
mechanism is explained. We also introduce
the setup of the paper and
recapitulate the eigenstates
(the wave functions) of the non-interacting 2+1 D Jackiw-Rebbi model.
In Sec.3, we investigate a toy model of QED$_3$ with a planar photon and calculate
the transition rate of the edge modes in the
framework of the Lindblad approach.
In Sec.4, the interaction with a real 3+1 D photon is studied.
Sec.5 provides the conclusion and outlook.

\section{Callan-Harvey mechanism}

In this section, we introduce the Callan-Harvey mechanism \cite{Callan_1985} in 2+1 D Jackiw-Rebbi model\cite{Jackiw&Rebbi_1976},
and the eigen-states of the non-interacting theory to lay the foundation of the next
sections.

In order to explain the Callan-Harvey mechanism, we start with a quite general 2+1 D fermion (electron) $\psi$
in the background of an Abelian gauge field $A_{\mu}$ ($\mu = 0\sim 2$) with the action
\begin{eqnarray}\label{action_1}
S_1
= \int d^3 z
  \bar{\psi} (\gamma^0 iD_0+ v \gamma^j iD_j -mv^2)\psi
\end{eqnarray}
in which $z^{\mu}= (t, x,y)$ with $\mu \in \{0,1,2\}$, $D_{\mu}= \partial_{\mu}-ie A_{\mu}$
and $j \in \{1,2\}$.
We assume $\mu=0$ is for the time component and $\mu=1,2$ or $j$ for the spatial components.
The Dirac matrices $\gamma^{\mu}$'s are given by
$\gamma^0=\sigma_z$,
$\gamma^1=i\sigma_y$ and $\gamma^2=-i\sigma_x$;
$\bar{\psi} = \psi^{\dagger} \gamma_0$,
 $v$ is the velocity of the fermions,
and the mass term $m=m(x)$ has the following domain wall structure
\begin{eqnarray}\label{mass}
m(x)=\left\{
\begin{aligned}
&& m_0 \quad x>0 \\
&& -M \quad x<0.
\end{aligned}
\right.
\end{eqnarray}
%
The mass parameters here $m_0$ and $M$ are positive. The action Eq.\ref{action_1}
with the domain-wall mass is
called Jackiw-Rebbi model \cite{Jackiw&Rebbi_1976}. In the original work
of Jackiw and Rebbi, they considered the special case when $m_0=M$.

Callan and Harvey considered
the effective Chern-Simons action of such a fermion theory  with a domain wall mass.
The Chern-Simons action can be obtained by integrating out the fermions and
its form is given by \cite{Callan_1985}
\begin{eqnarray}\label{Chern-Simons}
S_{CS}
=\frac{e^2}{h} \int d^3 x \,\mathcal{C}\, \epsilon^{\mu\nu\rho}A_{\mu}\partial_{\nu}A_{\rho}
\end{eqnarray}
with the Chern number $\mathcal{C}= {\rm sgn}(m(x))/2$.
This effective action varies by a boundary term under gauge transformations
in the presence of the domain wall.
This is an example of a gauge anomaly.
Fortunately, a zero mode living on the domain wall was found to
produce a chiral anomaly, which precisely cancels the aforementioned gauge anomaly.
In this sense, the bulk and the boundary exist in mutual dependence of each other.
Later in the 1990s, Chandrasekharan proved explicitly
such a cancellation \cite{Chandrasekharan_1994}.

\begin{figure}
  \includegraphics[width=0.3\linewidth]{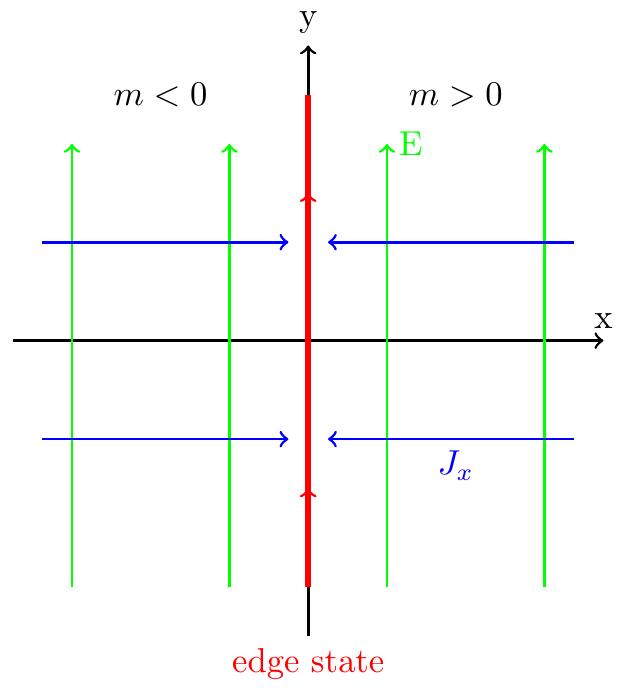}
  \caption{The setup of the system.
As explained in the main text of Sec.1, the fermion mass changes sign at $x=0$, and
therefore the Chern number is $1/2$ at the $x>0$ region and $-1/2$ at the $x<0$ region.
In the presence of an uniform electric field in the y-direction (green arrows),
the Hall currents (blue arrows) in
the two regions flow in opposite directions.
Therefore, there will be charge accumulation
at the edge.
  The red arrows denote the edge current.}
  \label{fig:setup}
\end{figure}

This anomaly cancellation can also  be  understood
at the level of the fermionic theory
i.e. before integrating the fermions to obtain
the effective Chern-Simons theory, Eq.\ref{Chern-Simons}.
Consider a 2+1 D Dirac fermion with constant mass term $m$.
The coupling of the fermion to the gauge field 
induces the parity anomaly in the electric current
\cite{Semenoff_1984}
\begin{eqnarray}\label{parity_anomaly}
e j_{\mu}
=\mathcal{C} \frac{e^2}{h}  \epsilon^{\mu\nu\rho}\partial_{\nu}A_{\rho},
\end{eqnarray}
with $\mathcal{C} = {\rm sgn}(m)/2$.
If the mass term is given by Eq.\ref{mass},
then the Hall conductivity
$\sigma_H = sgn(m)e^2/2h$,
changes its sign from the $x>0$ region to
the $x<0$ region.
Now we apply an electric field in the $y$-direction,
which induces Hall bulk currents in the $x$-direction.
Due to the sign change of the fermion mass, the Hall
currents in $x<0$ region and $x>0$ region flow in opposite
directions (See Fig.\ref{fig:setup}).
It leads to charge accumulation at the
edge region $x\sim 0$.
These currents are called Goldstone-Wilczek currents \cite{Chandrasekharan_1994}
or also anomaly inflow \cite{Xiong_2013, Fukushima_2018}.
If one neglects the edge mode, the fermions seem to disappear at the boundary which breaks the charge conservation and
leads to a gauge anomaly.

\begin{figure}
  \begin{subfigure}{0.3\linewidth}
  \includegraphics[width=\linewidth]{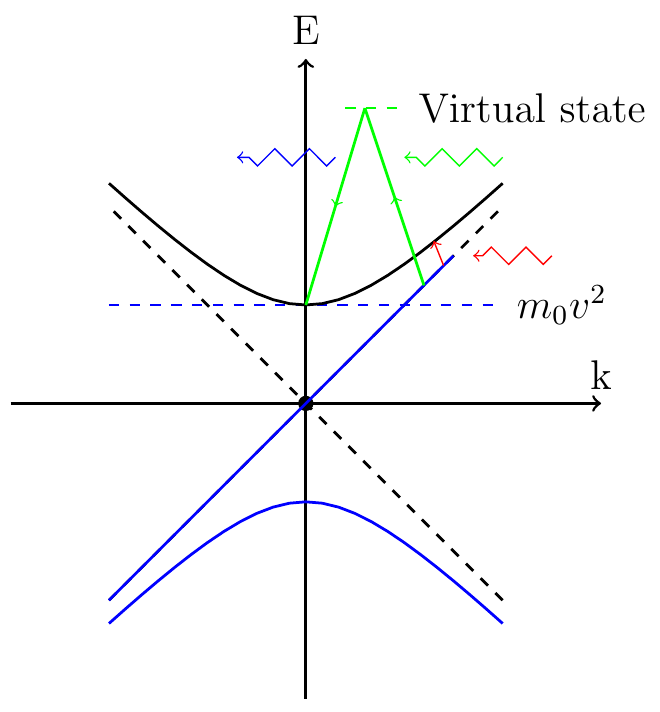}
  \caption{}
  \label{fig:energy_level_1}
  \end{subfigure}\\   
  \begin{subfigure}{0.3\linewidth}
  \includegraphics[width=\linewidth]{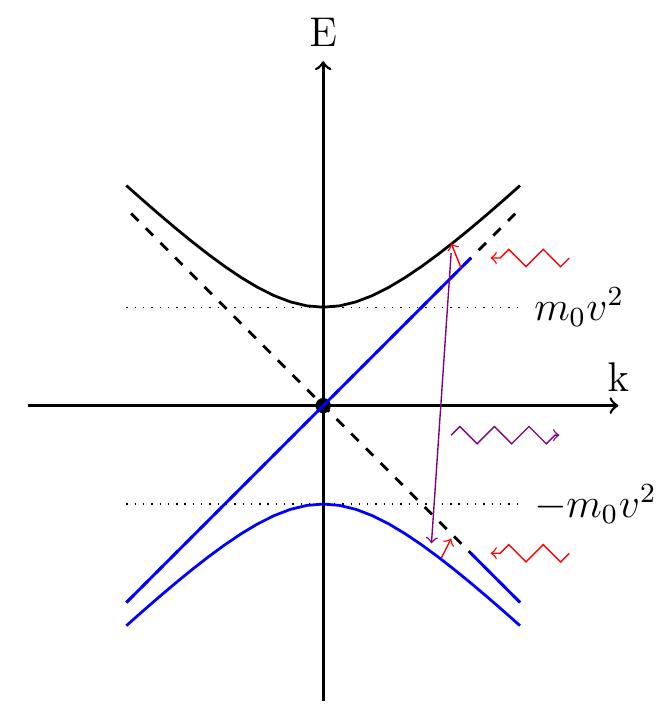}
  \caption{}
  \label{fig:energy_level_2}
  \end{subfigure}
\caption{The energy spectra for electron systems
and the edge-to-bulk transition processes. 
The blue and black lines describe the unoccupied and occupied states, respectively.
Zigzag lines represent ingoing or outgoing photons.
(a) A semi-infinite electron system with one edge or domain wall. There is
only one chiral mode, whose occupied states are depicted by the straight blue line.
 The red arrow depicts the leading-order excitation process,
which absorbs one photon. The green arrows are related to the second-order relaxation process,
which includes an absorption of a low-energy photon (green zigzag line) and an emission of
a high-energy photon (blue zigzag line).
(b) The schematic diagram of transitions in a finite-sized system with two edges.
The two straight blue lines represent the occupied edge states. The red and purple
arrows denote the absorption and emission processes, while the red and purple zigzag lines
denote the ingoing and outgoing photons, respectively.}
\end{figure}

One can also take the viewpoint of the edge states. According to
the so-called "bulk-edge correspondence", the fact that the difference
in Chern number between the
two sides of the domain wall
$\Delta \mathcal{C}= \mathcal{C}(x>0)-\mathcal{C}(x<0)
=\frac{1}{2} - (-\frac{1}{2})=  1$,
implies that there is one (massless) chiral  mode along the edge.
If one considers the interface between a Chern insulator with
$\mathcal{C}=1$ and the vacuum ($\mathcal{C}=0$),
then the difference in Chern number
is still 1, which also implies one chiral edge mode.
The main results of the two cases are essentially the same.
The dispersion relations for both bulk and edge modes
are shown in Fig.\ref{fig:energy_level_1}.
The chiral mode is described by 1+1 D massless Dirac equation,
and the chiral anomaly in 1+1 D tells us
$\partial_{\mu} j^{\mu}=\partial_{\mu} j_5^{\mu}= e E/h$.
(The first equality holds because there is only one
chiral mode, left-handed or right-handed.) If one looks at the
edge theory itself, the charge conservation is broken: the charge number
may increase with time \cite{Callan_1985}. Therefore,
one has to consider the edge and the bulk theory as a whole, and then
one will find that the total charge of the whole system is conserved:
the bulk loses charge, and the edge (the domain wall) gains the same amount of charge in turn.

However,
what is the fate
of this extra charge?
A similar charge pumping process was studied before
in the context of the QAHE under out-of-plane magnetic fields
\cite{Boettcher_2019, Tutschku_2020},
but the relaxation or dissipation process was not taken into account.
One expects some kinds of relaxation or transition process
which transfers the surplus electrons at the edge to the
bulk (See Fig.\ref{fig:energy_level_1} ).
Such a relaxation process can be mediated by an electron-boson coupling,
for example, via a photon or phonon \cite{Ulybyshev_2016}.
In the present work, such electron-photon interactions are investigated.
Since the speed of light $c$ is much bigger than the Fermi velocity $v$, i.e. $c>>v$,
the edge-state electron can be excited into bulk states via absorbing one photon,
at leading order in perturbation theory
(see the red arrow in Fig.\ref{fig:energy_level_1}).
At next-to-leading order, the edge state electron can absorb one photon first
and then emit another photon. Such a Compton scattering or Raman process
may also transfer the fermion from an edge state to the bulk
(see the green arrows in Fig.\ref{fig:energy_level_1}).

In the following, we construct the eigenstates (wave functions)
of the non-interacting theory,  i.e. $e=0$ in Eq.\ref{action_1}
and from now on we always assume
$0<m_0<< M$, i.e. the vacuum gap is much bigger than 
the massive Chern insulator gap.
Equivalently, the theory can be described in terms of the Hamiltonian
\begin{eqnarray}\label{Hamiltonian}
H
=     -vi \partial_x \sigma_x -vi \partial_y \sigma_y + m(x)v^2 \sigma_z.
\end{eqnarray}
The spinor $\psi$ has two components,
thus the equation $H\psi = E \psi$ includes two coupled first-order differential equations.

Since there is no $y$-dependence in the Hamiltonian \eqref{Hamiltonian}, the momentum along the $y$-direction is conserved,
such that the partial derivative $-i\partial_y$
can be replaced by a constant $p_2$.
Therefore, we assume $\psi(t,x,y)=\Xi(x) e^{-iEt + i p_2 y}$,
and then the spinor
$\Xi= (\xi_1, \xi_2)^T$
satisfies the following equation for $x>0$
\begin{equation}
\begin{pmatrix}
m_0 v^2 & -vi\partial_x-ivp_2  \\
-vi\partial_x+ivp_2  & -m_0 v^2
\end{pmatrix}
\begin{pmatrix}
\xi_1   \\
\xi_2
\end{pmatrix}
=E \begin{pmatrix}
\xi_1   \\
\xi_2
\end{pmatrix}.
\end{equation}
In order to solve the above differential equations, we
transform them into a second order differential
equation for component $\xi_1$:

\begin{equation}
\partial_x^2 \xi_1 +[(E/v)^2 -(m_0 v)^2 -p_2^2]\xi_1
=0.
\end{equation}
The other component $\xi_2$ can be expressed by $\xi_1$
as
\begin{equation}
\xi_2= \frac{(-iv\partial_x+ivp_2) \xi_1 }{E+(m_0 v)^2}.
\end{equation}
There is one edge state (bound state) localized around
$x=0$, which is given by
\begin{equation}
\Xi^{(e)} (x)=\sqrt{m_0 v}
\begin{pmatrix}
1   \\
i
\end{pmatrix} e^{-m_0 v x}
\end{equation}
with the energy $E=vp_2$. The bulk states (continuous states) are
\begin{equation}
\Xi_{p_1,p_2} (x)=a
\begin{pmatrix}
1   \\
\frac{vp_1+ivp_2}{E+m_0 v^2}
\end{pmatrix} e^{ip_1 x}
+b
\begin{pmatrix}
1   \\
\frac{-vp_1+ivp_2}{E+m_0 v^2}
\end{pmatrix} e^{-ip_1 x}
\end{equation}
The coefficients $a$ and $b$ are normalization constants, and can be found by the
normalization condition $\int \Xi_{p_1,p_2}^{\dagger}(x)\Xi_{p'_1,p_2}(x)dx= \delta(p_1-p'_1)$.
The result is given by
\begin{eqnarray}
    a=-\frac{E+m_0 v^2 -vp_2 - ivp_1}{2\sqrt{2\pi E(E-vp_2)}}
    \qquad
    \text{and}
    \qquad 
    b=\frac{E+m_0 v^2 -vp_2 + ivp_1}{2\sqrt{2\pi E(E-vp_2)}}
\end{eqnarray}

\section {Interaction with a planar photon (QED$_3$)}

In this section, we consider a 2+1 D Dirac fermion interacting
with a 2+1 D photon.
It is a toy model for the interaction between the electrons in a two-dimensional plane
and the photons.
While electrons can be confined to the two-dimensional plane in
the laboratory, photons only exist in three-dimensional space.
3+1 D photons will be considered in the next section.
It is instructive, however, to start with QED$_3$,
i.e. the case where both electrons and photons live in the two-dimensional plane.  
Furthermore, we take into account that the Fermi velocity for electrons in solids ís much smaller than the speed of light i.e. $v<<c$.

The action is given by
\begin{eqnarray}
S_2=S_1
+\frac{1}{2} \int d^3 z \Big(\dot{A_0}^2-c^2\sum_{i=1,2}(\partial_{i}A_0)^2\Big)
\end{eqnarray}
where $S_1$ is given in Eq.\ref{action_1} and
$A_0$ is the temporal component of the photon field.
The spatial components $A_1$ and $A_2$ are neglected in the following, because of the
small Fermi velocity \cite{Araki_2010}.
The interaction term $e\psi^{\dagger}\psi A_0$
is responsible for the transitions from edge states to
bulk states, and the coupling strength $e$
is the electron charge in 2+1 D, which has the dimension 1/2,
and scales as $E^{1/2}$, where $E$ is energy.

Since the speed of light is much bigger than
the Fermi velocity, $c>>v$, a transition
process from an edge state to
a upper-band bulk state with lower energy cannot happen at the first order,
i.e. a high-energy edge state
cannot decay into a low-energy upper-band bulk state, by emitting only one photon.
On the contrary, an electron at edge state can
absorb one photon and be excited into a bulk state with a higher energy
(see the red arrow in Fig. \ref{fig:energy_level_1} ).
Due to the photon absorption, a finite (nonzero) temperature
is necessary for such a process to occur.
This is the main focus of this section.
The leading order contribution to a real relaxation process
(from a high energy initial state to a low energy final state) comes from
the second order, which is similar to Compton scattering in
quantum electrodynamics.
It is depicted by the green arrows, in Fig. \ref{fig:energy_level_1}.
The corresponding process can be
described by the effective Hamiltonian
$H_{eff}=\lambda \psi^{\dagger}\psi A_0^2 $, where
$\lambda= e^2/E^*$, and $E^*$ is a characteristic energy related
to the virtual intermediate state (shown by the green dashed line
in Fig.\ref{fig:energy_level_1}). However, it is a high-order
process suppressed by the higher power of the coupling
constant.

The time evolution of the density matrix $\rho$ is governed by 
the equation
$\dot{\rho}=-i[H_I, \rho]$, where $H_I=e\psi^{\dagger}\psi A_0 $
is the interaction term of the
Hamiltonian in the interaction picture.
In Born approximation,
the total density matrix $\rho$ is assumed to be factorized into
$\rho= \rho_S \otimes \rho_B$, where $\rho_S$ is the
density matrix of the electron system and $\rho_B$ is
the density matrix of the bath or the bosons (photons).
Tracing out the
degree of freedom of the bath environment,
the evolution of the electron system $\rho_S=Tr_B(\rho)$
can be formulated by \cite{book_Breuer}
\begin{eqnarray}
\dot{\rho}_S
=-\int_0^t ds Tr_B [H_I(t),[H_I(s),\rho(s)]] \\
=-\int_0^{\infty} ds Tr_B [H_I(t),[H_I(t-s),\rho_S(t)\otimes\rho_B]].
\end{eqnarray}
in which the Markov approximation has been applied in the second equation.
We only consider the first order contribution in
perturbation theory, and assume that multi-particle excitations are suppressed.
Taking the average value on the edge state $|k_e>$ (the state
means adding one edge state
to the Fermi sea $|k_e>\otimes |FS> $,
but the Fermi sea $|FS> $ will not be mentioned below for simplicity),
one obtains the time evolution of the occupation probability
of the state $|k_e>$, which is given by $< k_e| \dot{\rho}_S |k_e>= -I_1+I_2 +h.c.$
with
\begin{eqnarray}\label{I1}
I_1=\int_0^{\infty} ds Tr_B < k_e| H_I(t) H_I(t-s) \rho_S(t)\otimes\rho_B |k_e>,
\end{eqnarray}
and
\begin{eqnarray}\label{I2}
I_2= \int_0^{\infty} ds Tr_B < k_e| H_I(t)\rho_S(t)\otimes\rho_B H_I(t-s)  |k_e>.
\end{eqnarray}
$I_1$ is the rate of the electron leaving from the edge state $|k_e>$ to bulk states,
while
$I_2$ is the rate of the electron coming to the state $|k_e>$.
These rates are related to the photon number distribution law.
The rate or
the speed of the latter process (photon emission) is
higher than the former one (photon absorption). Furthermore, at
exact zero temperature, the photon absorption process can not
happen at all, but the emission process can still happen.

In the present work, we consider nonzero temperature $T$ only in the photon sector of the theory,
such that the related thermal energy is much smaller than the bulk gap
$k_B T<< 2 m_0 v^2 $. Therefore, if the chemical potential of the edge state
$\mu << m_0 v^2 $, the temperature is not large enough to efficiently supply a photon for the excitation of edge state into a bulk one.
On the other hand, if edge's $\mu > m_0 v^2 $ 
and we consider the edge state with momentum $k_e$ such that $v\sqrt{k_e^2 +(m_0 v)^2 }-vk_e \sim k_B T$,
the excitation process to the bulk can indeed happen, even at small temperature of the photon bath $k_B T<< 2 m_0 v^2$. We consider this process as a main contribution to the relaxation of the edge states, neglecting other possible processes. As was mentioned above, the "Compton-like" relaxation  depicted by the green arrows in Fig. \ref{fig:energy_level_1} is suppressed by the second power of the interaction constant. Furthermore,  we neglect the backward relaxation $I_2$. 

In order to devise the arguments in favor of this approximation, we consider a finite-width system, e.g. a ribbon, with two edges (domain-walls). Fig.\ref{fig:energy_level_2} shows the energy occupation state
and transition processes for the two-edge system. There are two edges states now denoted by the straight dashed lines and straight blue lines in Fig.\ref{fig:energy_level_2}. If the electric field is parallel to the edges,
the Hall current is perpendicular to the edges and drives the charge from one edge to another.
Therefore, the chemical potential of one edge will decrease (depletion process), and the chemical potential of the other edge will increase (accumulation process). Because the two edges are far from each other, direct transition from one edge to another is difficult, if not completely impossible. Direct calculation of the transition rate from edge to edge gives the estimation of the order of ${\rm exp}(-m_0 L_x)$, with $L_x$ the distance between the two edges, i.e. the width of the ribbon, and $m_0$ is the fermion mass in the bulk. In contrast, the transition rate from edge to bulk is of the order of $1/\sqrt{m_0 L_x}$. If $L_x$ is large enough ($m_0 L_x >> 1$), both rates are small, but the former is much smaller than the latter, thus we neglect the direct transitions from one edge to another. The $I_1$ relaxation processes happen in both edges, leading to the appearance of holes in the lower band in the bulk. It opens the possibility for the direct transitions from upper to lower band in the bulk  via the photon emission depicted by the purple arrows in Fig. \ref{fig:energy_level_2}. These processes are of the order of $1$. It means that they are much faster than all edge-bulk transitions and they keep the upper band of the bulk almost empty, thus suppressing the inverse bulk-to-edge transitions denoted by $I_2$.  Thus we conclude that the time of the whole edge-to-bulk relaxation is determined by the comparatively slower process $I_1$ showed by red arrows in Fig.\ref{fig:energy_level_2}.

In order to further calculate $I_1$, we neglect the off-diagonal
elements of the density matrix and
insert a complete set of states between the two $H_I$ operators
(many-particle excitations are neglected).
Then the rate $I_1$ can be reformulated into
\begin{eqnarray}\label{mater_equ}
I_1
&=&\int_0^{\infty} ds Tr_B < k_e| H_I(t) |\mathbf{p}> \frac{d^2 p}{(2\pi)^2} <\mathbf{p}|H_I(t-s) \rho_S(t)\otimes\rho_B |k_e>, \nonumber \\
&=&e^2 \int \frac{d^2 p}{(2\pi)^2} W_{k_e}(\mathbf{p})\, r(k_e,t),
\end{eqnarray}
where $|k_e>$ is the edge state with momentum $k_e$,
$|\mathbf{p}>$ is the bulk state with momentum $ \mathbf{p}=(p_1,p_2)$, and
the function $r(k_e,t)$ is defined by
$ <k'_e|\rho_S(t) |k_e>=r(k_e,t)\delta(k'_e-k_e)$.
The quantity
\begin{eqnarray}\label{W1}
W_{k_e}(\mathbf{p})&=& \int_0^{+\infty} ds \int_0^{+\infty} dx \int_0^{+\infty} dx'
                \int\int dy dy'
                 e^{iEs} \nonumber \\
            && f_{\mathbf{p}}(x) f^{*}_{\mathbf{p}}(x')
               e^{i(p_2-k_e)(y-y')} G(s,x-x',y-y')
\end{eqnarray}
where $f_{\mathbf{p}}(x)=\Xi^{(e)\dagger}\Xi_{\mathbf{p}}(x)$ is the inner product
of the spinors,  and $G(s,x-x',y-y')= Tr_B[\rho_B \phi_{-}(s,x-x',y-y')\phi_{+}(0,0,0) ]$
is the correlation function of the photon field. The field $A_0$
is decomposed  into $A_0 =\phi_{+}+\phi_{-}$, with $\phi_{+}$ the
positive frequency component including the annihilation operators  and
$\phi_{-}$ the negative frequency component including the creation operators.
The other combination $Tr_B[\rho_B \phi_{+}\phi_{-}]$ is neglected by virtue of
the rotating wave approximation \cite{book_Breuer}.

\begin{figure}
  \includegraphics[width=0.5\linewidth]{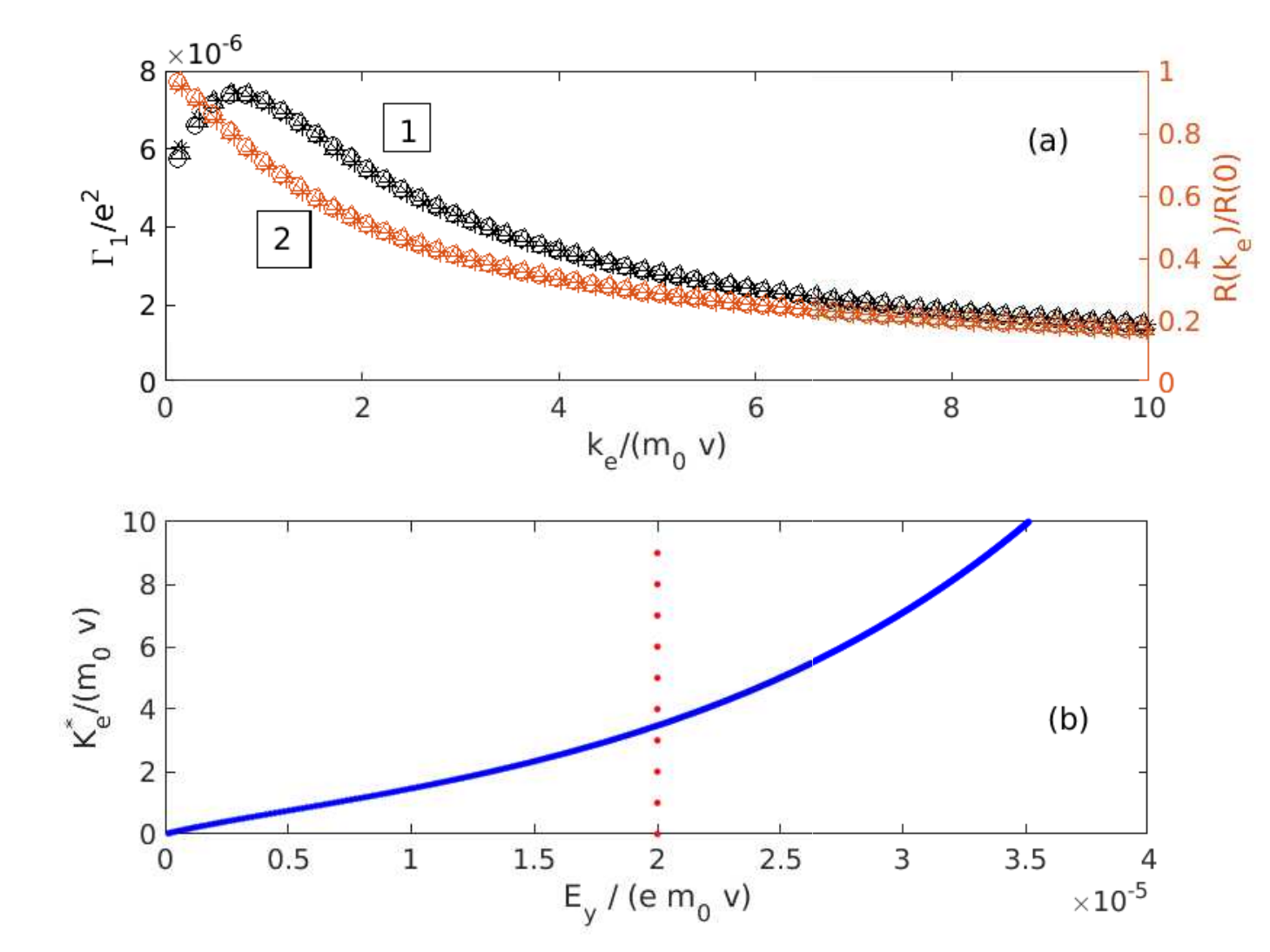}
  \caption{(a)The black curve (marked by label 1) shows the momentum
  dependence of the excitation rate $\Gamma_1(k_e)/e^2$ of the edge state $|k_e >$.
  It is computed according to Eq.\ref{Gamma_2}, with $v/c=0.01$.
  The horizontal axis is $k_e /(m_0 v)$. The black curve (marked by label 2) is the
  the distribution function $R(k_e)/R(0)$ in the saturation state,
  i.e. Eq.\ref{distribute}, with the electric field $E_y= 2\times 10^{-5} e m_0 v$
  related to the red dotted line in Fig.(b).
  (b) The saturation momentum $K_e^*$ obtained from Eq. \ref{K_star}
   as a function of electric filed $E_y$,
   in the step-function approximation for the
   edge-state occupancy.}
  \label{fig:transition_1}
\end{figure}

The photon correlation function $G$ can be calculated by mode expansion,
and the result (in Gaussian units) is
\begin{eqnarray}\label{Green_function}
G(t,x,y)= \int \frac{c^2 d^2 q}{(2\pi)^2 2 \omega_q} n_B(\omega_q)  e^{i\omega_q t-iq\cdot r},
\end{eqnarray}
where $n_B(\omega)=1/(e^{\beta \omega}-1)$ is the Bose-Einstein distribution function for the photon bath,
$\beta=1/({\rm k_B} T)$,
$\omega_q= c\sqrt{q_1^2+q_2^2}$, $q=(q_1, q_2)$
and $r=(x,y)$.
Therefore, we found
\begin{eqnarray}\label{W2}
W_{k_e}(\mathbf{p})=\int \frac{d^2 q}{(2\pi)^2 \omega_q}|F_{\mathbf{p},q_1}|^2 L_y
            \delta(p_2-k_e+q_2) \delta(E(k_e)-E_{\mathbf{p}}+\omega_q)n_B(\omega_q),
\end{eqnarray}
where $F_{\mathbf{p},q_1} = \int_0^{+\infty}  f_{\mathbf{p}}(x)  e^{iq_1 x} dx $ and
\begin{eqnarray}\label{F}
|F_{\mathbf{p},q_1}|^2 = \frac{v^2 p_1 q_1}{\pi E_{\mathbf{p}}(E_{\mathbf{p}}-vp_2)}
\Big[\frac{m_0 v}{(q_1-p_1)^2+(m_0 v)^2}  - \frac{m_0 v}{(q_1+p_1)^2+(m_0 v)^2}   \Big].
\end{eqnarray}
For simplicity, the function $\frac{m_0 v}{q^2+(m_0 v)^2}$ is replaced by
$\pi  \delta(q)$, and then function $W_{k_e}(\mathbf{p})$ can be evaluated as
\begin{eqnarray}\label{W3}
W_{k_e}(\mathbf{p})= L_y  \delta(E(k_e)-E_{\mathbf{p}}+\omega_{p_1,k_e-p_2})
              \frac{v^2 p_1^2 n_B(\omega_{p_1,k_e-p_2})}{\pi E_{\mathbf{p}}
              (E_{\mathbf{p}}-vp_2)\omega_{p_1,k_e-p_2}}.
\end{eqnarray}
Therefore, the transition rate of the edge state $|k_e>$ to the
bulk states is given by
\begin{eqnarray}\label{Gamma_1}
\Gamma(k_e)= e^2 \int W_{k_e}(\mathbf{p})\frac{d^2 p}{(2\pi)^2}.
\end{eqnarray}
Its integrand includes the delta function $\delta(E(k_e)-E_{\mathbf{p}}+\omega_{p_1,k_e-k})$,
with $E_{p_1,p_2}= v \sqrt{p_1^2+p_2^2 + (m_0 v)^2}$ and
$\omega_{p_1,p_2}=c \sqrt{p_1^2+p_2^2}$. If the Fermi velocity $v$
is much smaller than the speed of light $c$, i.e. $v/c \sim 1/100$,
then it is safe and convenient to replace $E_{\mathbf{p}}$ in the integrand
by $E_{0,k_e}$. After integrations, we obtain the result of the transition
rate per unit length $\Gamma_1 (k_e)= \Gamma(k_e)/L_y$ as follows
\begin{eqnarray}\label{Gamma_2}
\Gamma_1(k_e)=  \frac{ e^2 v^2\Delta E}{\pi c^2 E_{0,k_e}} n_B(\Delta E),
\end{eqnarray}
with $\Delta E = E_{0,k_e} - E(k_e)$. When $k_e \rightarrow +\infty$,
$\Delta E \rightarrow 0^+$ and $\Gamma_1(k_e)$ goes to zero as $\sim {\rm k_B} T/k_e$.
The $k_e$-dependence of function $\Gamma_1(k_e)$ is shown by curve 1
in Fig.\ref{fig:transition_1}(a).

Now let us analyze the consequences of such an excitation process,
and consider the evolution of the occupancy of the
edge states. Suppose at time $t=0$, the chemical potential
of the whole system is at $\mu=0$ (Fig.\ref{fig:energy_level_1}), and one turns on
the electric field in the y-direction $E_y$.
On the one hand, because of the electric field,
there is a constant rate of electrons flowing toward the edge and accumulating
there. On the other hand, the accumulated electrons at the edge
are excited via thermal fluctuations and transferred
to the bulk.
If we assume that at a time $t$ the edge states are occupied
up to the momentum $K_e(t)$ , what is its
behavior at the later times  $t\rightarrow \infty$?
Is it possible for the system to reach a saturation?
A "saturation" means  a balance between the inflow current
towards the edge and the excitation process depleting
the edge.  The excitation rate
from the edge to the bulk $\int_0^{K_e}\Gamma_1(k_e)dk_e$ is small in the beginning (for small $t$) because $K_e$ is small i.e. $K_e(t)\sim 0$.
Therefore the accumulation process is stronger
than the depletion, and $K_e$ starts to increase. When $K_e$ increases,
the rate of depletion also increases. If the depletion rate coincides with the
accumulation rate, the process reaches equilibrium, and $K_e$ saturates. In order to calculate the saturation momentum $K_e^{*}$, we equate the
two rates (number of particles per unit time and per unit length)
\begin{eqnarray}\label{K_star}
\int_0^{K_e^{*}}\Gamma_1(k_e)dk_e= \sigma_H E_y/e .
\end{eqnarray}
Finite values of $K_e^{*}$ can be found, as a function of $E_y$,
which is shown in Fig.\ref{fig:transition_1} (b). Asymptotically,
$K_e^{*}(E_y)$ scales as $\sim {\rm exp}(E_y/e m_0 v)$, for large $E_y$.

Above, we assumed the distribution function
on the edge $r(k_e)$ to be
a step function: $r(k_e)=1$
when $k_e < K_e$ and $r(k_e)=0$ when $k_e > K_e$.
Such an assumption is simple, but is not entirely realistic.
In order to approach reality, we lift such an assumption,
and allow the  distribution function $r(k_e,t)$ to take any value between
zero and one. We are going to find such a distribution function
at the saturation ($t\rightarrow +\infty$).
Suppose $\Delta t$ is a very short time interval and $k'_{e}= k_{e}+E\Delta t$,
and then we have $r(k'_{e}, t+\Delta t)=r(k_{e},t)(1-\Gamma_1 (k_e)\Delta t)$.
It means that the momentum of the edge-state fermions
is changed by the electric field $E_y$ during the time interval,
and in the meantime, the fermions leave the edge (via excitation process)
at the rate $\Gamma_1$.
In the stationary state, $r(k_{e},t+\Delta t)=r(k_{e},t)$, which
doesn't depend on time, and can be denoted by the function $R(k_e)$.
Therefore, we obtain $R'(k_e)E_y=-\Gamma_1 (k_e)R(k_e)$, from which
we find the function $R(k_e)$ as the final
distribution along the edge:
\begin{eqnarray}\label{distribute}
R(k_e)/R(0)=  {\rm exp} \Big( -\int_0^{k_e} \Gamma_1 (k')dk'/E_y  \Big),
\end{eqnarray}
which is shown by curve 2 in Fig.\ref{fig:transition_1} (a).

\section {Interaction with 3+1 D photons}

In this section we consider a  realistic model, where the 2+1 D electrons interact
with 3+1 D photons. The corresponding action is given by
\begin{eqnarray}
S_3=S_1+
\frac{1}{2} \int d^4 x \Big(\dot{A_0}^2-c^2\sum_{i=1}^{3}(\partial_{i}A_0)^2\Big),
\end{eqnarray}
where $d^4 x = dt\, dx\, dy\, dz$.
The 2+1 D electron system is located on the $z=0$ plane.

\begin{figure}
  \includegraphics[width=0.5\linewidth]{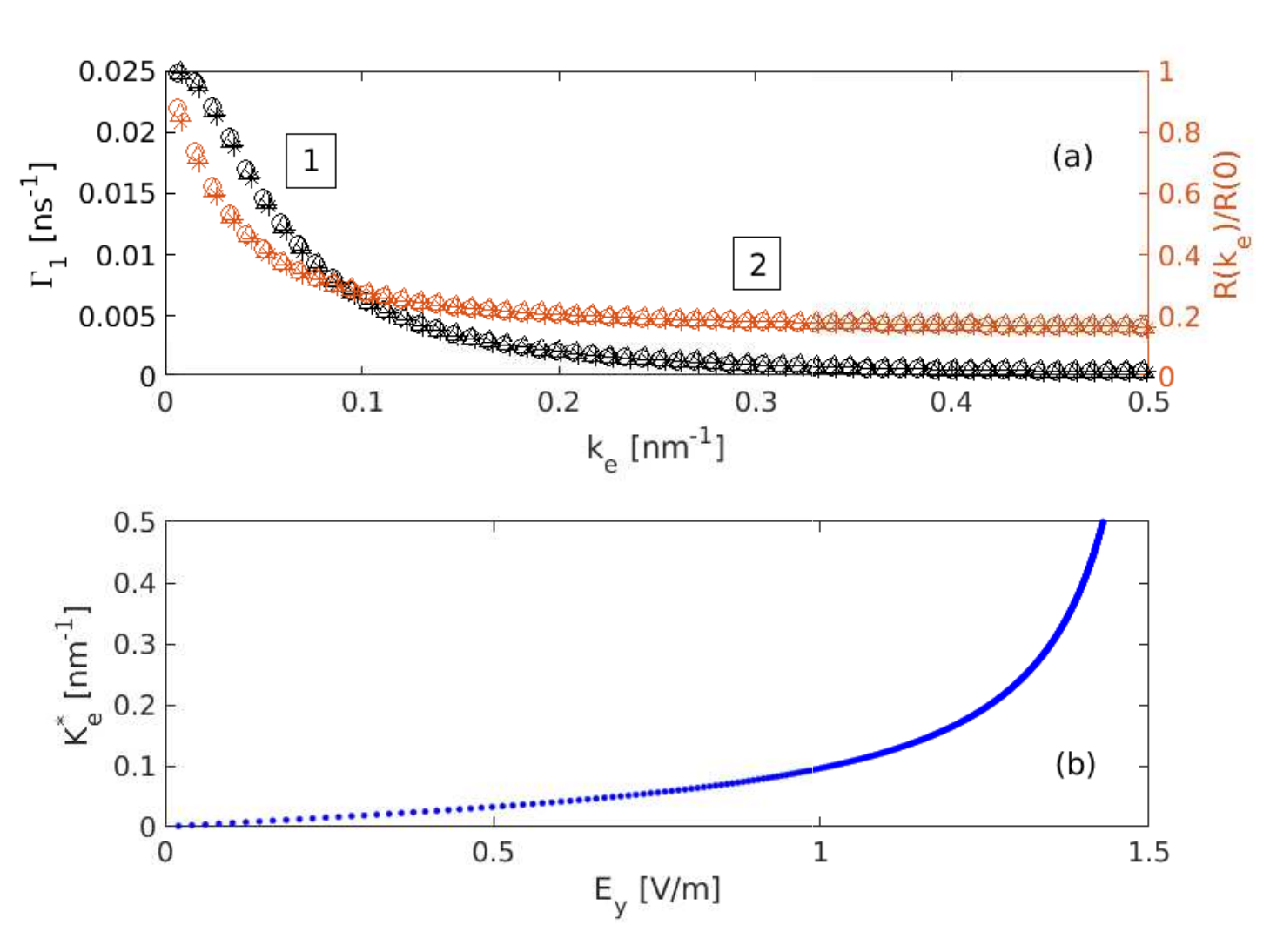}
  \caption{(a)The black curve (marked by label 1) is the excitation rate
  $\Gamma_1(k_e)$,
   as a function of momentum $k_e$, Eq.\ref{Gamma_3d}.
   ns is nanosecond.
   The parameters are given as follows:  the bulk gap 
   $\Delta= 2m_0 v^2=0.1 {\rm eV}$, Fermi velocity $v=0.01c$ 
   and the temperature ${\rm k_B} T=\Delta/4 $.
   The red curve (marked by label 2) is the
   the occupation function $R(k_e)/R(0)$ in the saturation state,
   with $E_y= 0.75 \, V/m $. 
   It is obtained from Eq.\ref{distribute}, with the $\Gamma_1(k_e)$ function given
   by Eq.\ref{Gamma_3d}.
   (b) The saturation momentum $K_e^*$ in the step-function assumption for the edge occupation,
   v.s. electric filed $E_y$. One observes that there is a critical electric field, around $1.5\, V/m$, 
   above which    the value of $K_e^*$ diverges.}
  \label{fig:transition_rate_3d}
\end{figure}

The deduction in the previous section about the evolution of the density
matrix and the transition rate can be repeated straightforwardly. However, the
photon correlation function $G$ in Eq.\ref{Green_function} has to be modified,
because of the
different dimensionality. As for 3+1 D photon, the corresponding correlation
function $\mathcal{G}(s,x-x',y-y')$ is defined as
\begin{eqnarray}
\mathcal{G}(s,x-x',y-y')= Tr_B[\rho_B \phi_{-}(s,x,y,0)\phi_{+}(0,x',y',0) ],
\end{eqnarray}
where the $z$ component of the spatial coordinates is fixed to be 0, because
the photons interact with the fermions only at the $z=0$ plane.
The result of $\mathcal{G}$ can be obtained  by mode expansion
\begin{eqnarray}\label{Green_function_3d}
\mathcal{G} (t,x,y)= \int \frac{d^3 q}{(2\pi)^3 2 \omega_q}
                     n_B(\omega_q)  e^{i\omega_q t-iq_1 x_1 -iq_2 x_2},
\end{eqnarray}
where $q=(q_1, q_2, q_3)$, and $\omega_q = c\sqrt{q_1^2+ q_2^2+ q_3^2} $.
Corresponding to Eq.\ref{W3},  the function $W$ for 3+1 D photon will
be given by
\begin{eqnarray}\label{W4}
W_{k_e}(\mathbf{p})=L_y \int \frac{dq_3}{2\pi}\delta(E(k_e)-E_{\mathbf{p}}+\omega_{p_1,k_e-p_2,q_3})
             \frac{v^2 p_1^2 n_B(\omega_{p_1,k_e-p_2,q_3})}{\pi E_{\mathbf{p}}
             (E_{\mathbf{p}}-vp_2)\omega_{p_1,k_e-p_2,q_3}}.
\end{eqnarray}
From the function $W_{k_e}(\mathbf{p})$, we obtained the transition rate
of the edge state $k_e$ to the bulk
$$\Gamma_1(k_e)=(e^2/L_y) \int W_{k_e}(\mathbf{p})\, d^2 p/(2\pi)^2,$$ and its result is given by
\begin{eqnarray}\label{Gamma_3d}
\Gamma_1(k_e)=  \frac{4 \alpha v^2\Delta E^2 n_B(\Delta E)}{3 c^2  E_{0,k_e} } .
\end{eqnarray}
with $\alpha= e^2/c$ and $\Delta E = E_{0,k_e}-vk_e$.
Then as in the previous section, one can figure out the charge
accumulation at the edge and the stationary distribution law of
the edge electrons in momentum space, which is shown in
Fig.\ref{fig:transition_rate_3d}.

  From Eq.\ref{Gamma_3d}, one notices that when $k_e \rightarrow +\infty$,
$\Gamma_1(k_e)$ goes to zero as $1/k^2_e$, implying that
$\int^{+\infty} \Gamma_1(k_e) dk_e$ is a finite number.
As in the previous section,
the saturation momentum $K_e^{*}$ can be specified by
Eq. \ref{K_star} according to the assumption
of step-function edge-state distribution.
However, if the electric field $E_y$ is larger than
the critical field $E_y^{(c)}=
\int_0^{+\infty}\Gamma_1(k_e)dk_e/(\sigma_H L_y)$,
then the saturation momentum $K_e^{*}$ will be
infinite. It means all the edge states will be occupied,
if the electric field is strong enough.
This "electron avalanche" phenomenon is due to
our low-energy effective model 
which is not regularized by the high energy part of the dispersion relation  as it always appears in real materials.
If $K_e^{*}$ is very large,
the higher momentum part of the  band structure should be taken into account, 
and the dispersion curve will bend, which prevents
$K_e^{*}$ from going to infinity.

We will now discuss the realization of such an effect
in the laboratory,
and estimate the order of magnitudes for the physical quantities.
In the last section, we considered
half infinite planar systems,
and infinitely long ribbon-shaped systems.
The former one has only one boundary, while
the latter one has two boundaries.
However, both of them
are hard to realize in experiment.
Instead of these infinite-sized systems,
we consider a cylinder with finite length or an annulus as more realistic examples. Both of them are finite sized and have one hole and two edges.
If the magnetic field going through the hollow part of
this kind of a system varies with time, the 
electric field parallel to the edges appears automatically.
Due to this electric field,
the Hall current is perpendicular to the edges and drives
the charge from one edge to another.
Therefore, the chemical potential of one edge will
decrease (depletion process), and the chemical potential of the other edge
will increase (accumulation process),  exactly as shown in Fig.\ref{fig:energy_level_2}.

At last, we estimate the orders of the main quantities, such as the
transition rate $\Gamma_1(k_e)$ and the critical electric field $E_y^{(c)}$.
In Fig. \ref{fig:transition_rate_3d} we show how the edge-to-bulk
transition rate changes with the wave vector of the edge state $k_e$. 
One can see that the transition rate is of the order of ns$^{-1}$ (nanosecond) and 
it decreases with the increase of the wave vector $k_e$.
For a massive Chern insulator with gap $\Delta=0.1 {\rm eV}$, at temperature
given by ${\rm k_B} T=\Delta/4 $, i.e. $T\sim 250 \, {\rm K}$, the
rate at which edge-state electrons transition into the bulk (per unit length of the edge)
is about $3\times 10^{14}\, m^{-1} s^{-1}$. If the size of a sample
is $1 \mu m$, the rate is $3\times 10^{8}\, s^{-1}$.
It means the life time of an edge state will be about 3 $n s$.
The critical electric field is about 1.5 V/m, which does not
depend on the the size of the sample.

\section {Conclusion and Outlook}

In the present work, we revisited the Callan-Harvey mechanism
in Jackiw-Rebbi model with a space-dependent domain wall mass.
Due to the parity anomaly, the electric field, 
which is parallel to the domain wall (the edge), 
drives the electrons to the edge. 
As the electrons accumulate along the edge,
they starts to transfer into the bulk states via thermal fluctuation.
We studied the time evolution of the surplus charge at the edge
in the Lindblad formalism, and the transition rate from the edge to the bulk was calculated.
In such a transition process, photon absorption is necessary. Therefore,
at zero temperature, the transition process does not occur
in our electron-photon interaction model, and the charge accumulation
at the edge will be boundless.
At finite (non-zero) temperature,
we studied the stationary state at late times $t\rightarrow +\infty$.
In the planar photon (QED$_3$) case, the stationary state can be obtained
for arbitrary electric field. In the 3+1 D photon case,
there is a critical electric field strength $E^{(c)}$, below which
the stationary state exists, but above which the stationary state
does not exist and the charge accumulation will be boundless.

Our present study investigated the effects of electron-photon interaction
on the edge states, and improved the physical picture of 
Callan-Harvey mechanism with dissipation processes.
It has not only scholar interest from quantum field theories, but also
might have potential applications 
in the condensed matter (optical relaxation in topological materials) 
and potential applications in engineering.
For example, the optical processes depicted in Fig.\ref{fig:energy_level_2} might
make such a system into a new light source:
in the presence of an electric field, the system
absorbs two low-energy photons from the thermal bath (the environment), 
and then emits
one high-energy photon, with the energy $\sim \Delta$ the band gap.
Furthermore, if one replaces the (low-energy) thermal photons
(the red zigzag lines in Fig.\ref{fig:energy_level_2}) by
incident photons with the same energy, then the incident
photons trigger the relaxation (the purple zigzag line
in Fig.\ref{fig:energy_level_2}), and vice versa.

There are also several directions for the future.
In the present work, we considered the Dirac mass term, which
is a constant within a bulk region. A natural generalization is
to study the momentum dependent mass, as in the Bernevig–Hughes–Zhang (BHZ) model.
Besides, electron-phonon interactions should be taken into account  
in the condensed matter systems, 
and the heat dissipation effect can be studied.

\begin{acknowledgments}
We acknowledge funding by the Deutsche Forschungsgemeinschaft (DFG, German Research
Foundation) through SFB 1170, Project-ID 258499086, through the W\"{u}rzburg-Dresden Cluster
of Excellence on Complexity and Topology in Quantum Matter – ct.qmat (EXC2147, Project-
ID 390858490) as well as by the ENB Graduate School on Topological Insulators.
M.U.  thanks  the  DFG   for financial support  under the project UL444/2-1. 
C.N. thanks the support by the Israel Science Foundation
(grant No. 1417/21) and by the German Research Foundation through a German-Israeli Project Cooperation (DIP) grant “Holography and the Swampland” and by  Carole and Marcus Weinstein through the BGU Presidential Faculty Recruitment Fund.

\end{acknowledgments}


\end{document}